\documentclass[technote]{IEEEtran}

\author{Hao Chen\\
Software Engineering Institute\\
East China Normal University\\
Shanghai 200062, P.R. China\\
haochen@sei.ecnu.edu.cn}

\title{\bf CRT-Based High Speed Parallel Architecture for Long BCH Encoding }

\begin{document}

\maketitle
\begin{abstract}

BCH (Bose-Chaudhuri-Hocquenghen) error correcting codes ([1]-[2])
are now widely used in communication systems and digital technology.
Direct LFSR(linear feedback shifted register)-based encoding of a
long BCH code suffers from serial-in and serial-out limitation and
large fanout effect of some XOR gates. This makes the LFSR-based
encoders of long BCH codes cannot keep up with the data transmission
speed in some applications. Several parallel long parallel encoders
for long cyclic codes have been proposed in [3]-[8]. The technique
for eliminating the large fanout effect by J-unfolding method and
some algebraic manipulation was presented in [7] and [8] . In this
paper we propose a CRT(Chinese Remainder Theorem)-based parallel
architecture for long BCH encoding. Our novel technique can be used
to eliminate the fanout bottleneck. The only restriction on the
speed of long BCH encoding of our CRT-based architecture is
$log_2N$, where $N$ is the length of
the BCH code.\\


\end{abstract}

{\bf I. Introduction and Preliminaries}\\

BCH codes were introduced in [1-2] and have been extensively studied
. Let $GF(2^t)$ be a finite field of $2^t$ elements and $\alpha \in
GF(2^t)^*=GF(2^t)-\{0\}$ be a primitive element. $(c_0,...,c_{N-1})
\in GF(2)^N$, where, $N=2^t-1$, is a codeword of the BCH code
$C(\delta)$ of designed distance $\delta$, if $\Sigma_{i=0}^{N-1}
c_i \alpha^{ji}=0$ for $j=1,2,...,\delta-1$. It is well-known that
the minimum Hamming distance of $C(\delta)$ is at least $\delta$.
For any polynomial in $GF(2)[x]$, it can be factorized to the
product of some irreducible polynomials in $GF(2)[x]$(see [11]). Let
$w_1(x),...,w_r(x)\in GF(2)[x]$ be the distinct monic irreducible
polynomials whose zeros are of the form $\alpha^{2^dj} \in GF(2^t)$,
where $d$ is arbitrary non negative integer . We know that the
generator polynomial $g(x)$ is the product $g(x)=w_1(x)\cdots
w_r(x)$. It is clear
$deg(w_i(x)) \leq t$ for $i=1,...,r$ (see [10]).\\

{\bf Example 1(see [10]).} Let $C \subset GF(2)^{15}$ be the $[15,
5, \geq 7]$ BCH code with zeros $\alpha^1,...,\alpha^6 \in GF(16)$,
where $\alpha$ is a primitive element of $GF(16)$. Its generator
polynomial
$g(x)=x^{10}+x^8+x^5+x^4+x^2+x+1=(x^4+x+)(x^4+x^3+x^2+x+1)(x^2+x+1)$.\\

For a BCH code of length $2^t-1$ and dimension $2^t-1-deg(g(x))$,
the number  $r$ is determined by the cyclotomic coset decomposition
of the zero set $\{\alpha^1,...,\alpha^{\delta-1}\}$.\\

{\bf Example 1 (continued, see [10]).} We have 3 cyclotomic coset
$\{\alpha^1,\alpha^2,\alpha^4,\alpha^8,\}$,
$\{\alpha^3,\alpha^6,\alpha^12,\alpha^9\}$,
$\{\alpha^5,\alpha^{10}\}$. Thus we have $r=3$.\\

Since $1$ is not a zero of the polynomial $g(x)$, it is clear $ r <
\frac{deg(g(x)}{2}$. Actually, most of the cyclotomic sets are of
the size $t$, $r$ is roughly $\frac{deg(g(x))}{t}$. When $t$ is a
prime number, we know each cyclotomic coset except $\{1\}$ is of the
size $t$(see [11]). Thus $r=\frac{deg(g(x)}{t}$ when $t$ is a
prime(see Example 2 and 3 below).\\

The systemic encoding of a cyclic code with generator polynomial
$g(x)$ ($deg(g(x))=n-k$) and code length $n$ is processed as
follows. For a $k$-bit message ${\bf m}=(m_{k-1},...,m_0) \in
GF(2)^k$, set $m(x)=m_{k-1}x^{k-1}+ \cdots +m_1x+m_0 \in GF(2)[x]$,
then the encoded codeword is ${\bf c}=(c_{n-1},...,c_0) \in GF(2)^n$
such that $c(x)=c_{n-1}x^{n-1}+ \cdots +c_1x+c_0
=m(x)x^{n-k}+Rem_{g(x)}(m(x)x^{n-k})$, where $Rem_{g(x)}(f(x))$ is
the remainder polynomial dividing $f(x)$ by $g(x)$, that is,
$f(x)=q(x)g(x)+Rem_{g(x)}(f(x))$, $deg(Rem_{g(x)}(f(x) ))<
deg(g(x))$.\\

For multiplying the input polynomial $u(x) \in GF(2)[x]$  by a
polynomial $h(x) \in GF(2)[x]$, we have a LFSR circuit to implement
the multiplication with at most $nz(h)\leq deg(h(x))+1$ XOR gates,
where $nz(h)$ is the number of non zero coefficients in $h(x)$. For
dividing the input polynomial $u(x) \in GF(2)[x]$  by the polynomial
$h(x) \in GF(2)[x]$, we have a LFSR circuit with at most $nz(h)$ XOR
gates, which outputs the
remainder polynomial of the division (see [10-11]).\\

Long BCH codes can sometimes achieve better performance than
RS(Reed-Solomon) codes, which is now widely used in digital video
broadcasting, optical communication and magnetic recording systems.
Hence BCH codes are of great interest. Long BCH encoding and
decoding can be implemented directly by linear feedback shifted
register(see [3] and [10]). However this LFSR-based architecture
suffers from serial-in and serial-out limitation and large fanout
effect. The LFSR-based systemic encoding of a long BCH code is
actually a {\em division circuit} with the divisor $g(x)$ (see [10])
and the large fanout of some XOR gate would lead to large gate
delay.  In high-speed applications such as optical communication
systems and digital video broadcasting, such LFSR-based long BCH
encoding cannot keep up with the data
transmission speed. Thus faster parallel processing of long BCH encoding is needed.\\

Several parallel encoding architectures for long cyclic codes  have
been proposed in [4-6]. In [7] and [8], K. K. Parhi et al presented
the technique of parallel architecture of long BCH encoding based on
J-unfolding method( see [9]), which
can eliminate the large fanout effect. \\

In this paper we give a parallel architecture of long BCH encoding
which is based on Chinese Remainder Theorem (CRT, see [11]).  The
basic idea is  the transformation of the above long division LFSR
circuit of the generator polynomial $g(x)$ by several short
 division LFSR circuits of low degree polynomials $w_1(x),...,w_r(x)$
in parallel. In this process, we need some multiplication LFSR
circuits which have no large fanout. The advantage of our novel
parallel architecture is the only limitation on number of the fanout
of the CRT-based long BCH encoding is $log_2N$ , where $N$ the code
length of the BCH
code.\\

{\bf II. CRT-based Parallel Architecture of Long BCH Encoding}\\

We need to recall Chinese Remainder Theorem. Let $f(x), g(x) \in
GF(2)[x]$ be two polynomials. Suppose $g(x)=g_1(x) \cdots g_r(x)$,
where  $g_1(x),...,g_r(x)$ are pairwise co-prime, that is,
$gcd(g_i(x),g_j(x))=1$ for any two distinct $i$ and $j$. Let
$g_i'(x)=\frac{g(x)}{g_i(x)} \in GF(2)[x]$ for $i=1,...,r$. It is
clear $deg(g_i'(x))=deg(g(x))-deg(g_i(x))$ and
$gcd(g_i'(x),g_i(x))=1$. By using generalized Euclid algorithm we
can find a polynomial $g_i''(x)$ such that $deg(g_i''(x))
<deg(g_i(x))$ and $g_i''(x)g_i'(x) \equiv 1$ mod $g_i(x)$ (i.e.
$g_i''(x)g_i'(x)-1$ can be divided by $g_i(x)$). We have the following result.\\

{\bf CRT (see [11]).} {\em $Rem_{g(x)}(f(x))=\Sigma_{i=1}^r g_i'(x)
Rem_{g_i(x)}(g_i''(x)f(x))$.}\\

Let $g(x)=w_1(x) \cdots w_r(x)$ be the generator polynomial of a BCH
code, where $w_1(x),...,w_r(x)$ are the distinct irreducible
polynomials in $GF(2)[x]$ as in the previous section. From the
theory of the finite field(see [11]),  $deg(w_i(x)) \leq t$, a fixed
constant around $log_2N$, where $N=2^t-1$ is the code length of the
BCH code. It is clear that these polynomials are pairwise co-prime.
Set $w_i'(x)=\frac{g(x)}{w_i(x)}$. Let $u_i(x) \in GF(2)[x]$ be the
unique polynomial such that $u_i(x)w_i'(x) \equiv 1 $ $mod$
$w_i(x)$. \\

From CRT $Rem_{g(x)}(m(x)x^{n-k})=\Sigma_{i=1}^r
w_i'(x)Rem_{w_i(x)}(u_i(x)m(x)x^{n-k})$, we can have a parallel
architecture for getting $Rem_{g(x)}(m(x)x^{n-k})$ immediately.
First we have $r$ parallel LFSR circuits  multiplying
$u_1(x),...,u_r(x)$, then $r$ parallel LFSR circuits  dividing
$w_1(x),...,w_r(x)$;  $r$ parallel LFSR circuits  multiplying
$w_1'(x),...,w_r'(x)$ in the third step and finally a circuits
summing the outputs from the previous circuits. \\

Here the fanout effect of the LFSR circuits dividing by
$w_1(x),...,w_r(x)$ is upper bounded by $t$, which is around $logN$.
It is well known that the multiplying and summing LFSR circuits have
no large fanout effect and can be execute with small latency.
Comparing with the direct LFSR-based architecture , though the
number of clock cycles is perhaps increased in our architecture, the
clock period is substantially decreased by eliminating the large
fanout effect. Thus our parallel architecture of getting
$Rem_{g(x)}(m(x)x^{n-k})$ (the systemic encoding of the BCH code) is
suitable in the  high speed applications. The speed of this
CRT-based parallel architecture of long BCH encoding is essentially
dependent on the number $t$, which is around the $log_2 N$, where
$N$ is the code length
 of the BCH code.\\

{\bf III. Implementation and Further Comments}\\

In this section the implementation and the cost of the CRT-based
architecture of long BCH encoding are given.\\

{ \bf Implementation:}\\

 Step 1. Multiplication LFSR of polynomials
$u_1,...,u_r$ with the input polynomial $m(x)x^{n-k}$. Here the circuits need $\Sigma (deg(u_i)+1)$ XOR gates.\\

Step 2. Division LFSR of polynomials $w_1,...,w_r$ with the inputs
of outputs of the circuits in the Step 1. Here the
circuits need $\Sigma (deg(w_i)+1)$ XOR gates.\\

Step 3. Multiplication LFSR of polynomials $w_1',...,w_r'$ with the
inputs of the outputs of the circuits in the Step 2. Here the
circuit need $\Sigma(deg(g)-deg(w_i)+1)$ XOR gates.\\

Step 4. The summation LFSR of the $r$ outputs in Step 3. Here the
circuit need at most $r(t+1)$ XOR gates.\\

We can get a upper bound on the number of XOR gates used directly,
it is upper bounded by $\Sigma(
deg(u_i)+1+deg(w_i)+1+deg(w_i')+1)+r(t+1) \leq 2r(t+1)+r(deg(g)+2)$.
From the estimation of $r$, this number is roughly $2deg(g)+\frac{deg(g)}{t}(deg(g)+2)$.\\

 {\bf Example 2 (see [7]).} We consider the  BCH code with code length
 $2047=2^{11}-1$ and dimension $1926$. Its  generator polynomial is  a degree $121$
 polynomial $g(x) \in GF(2)[x]$, which is the product of $11$ distinct irreducible
 polynomials $w_1,...,w_{11}$ of degree 11. Thus our architecture need at most $1595$ XOR gates. The number of
 fanout is upper bounded by the $deg(w_1),...,deg(w_{11})$, which is at most 11. In some sense this is better then the
 architecture in [7].\\

{\bf Example 3 (see [8]).} We consider the BCH code with code length
$N=2^{13}-1$ and dimension $7684$. Its generator polynomial is a
degree $507$ polynomial $g(x)$ in $GF(2)[x]$. $g(x)$ is the product
of $39$ degree $13$ distinct irreducible polynomials in $GF(2)[x]$.
Thus our architecture need at most $20865$ XOR gates. The number of
fanout XOR gates in the architecture is at most $13$. In some aspect
this is better
than the architecture in [8].\\

It is clear the idea can be used for systematic encoding for any
 long cyclic code with generator polynomial $g(x)=g_1(x) \cdots
 g_r(x)$, where $g_1,...,g_r$ are pairwise co-prime polynomials. Secondly in
 some cases, if we can choose the generator polynomial with the same
 code parameters, it is better to use the generator polynomial $g(x)
 \in GF(2)[x]$ with the property that the numbers of nonzero coefficients in $g_1,...,g_r$
 are as small as possible. However the idea of CRT-based architecture can  not
be used  for the  encoding of long CRC codes (see [4-6]) because the
generator polynomials of CRC codes are
 irreducible.\\

{\bf IV. Conclusion}\\

In this paper we have presented a CRT-based high speed parallel
architecture for long BCH encoding. The architecture can be used to
eliminate the large fanout effect. The only limitation of this
CRT-based parallel architecture is the logarithm of the code length
of the BCH code. It should be noted that our architecture of using
CRT for transforming the long division LFSR of polynomial
$g(x)=g_1(x) \cdots g_r(x)$, where $g_1,...,g_r$ are pairwise
co-prime,  to short division LFSR in parallel can be used for
systematic encoding
of any long cyclic code generated by $g(x) \in GF(2)[x]$. \\

{\bf Acknowledgement:} The work was supported by the National
Natural Science Foundation of China Grant 90607005 and
60433050.\\

\begin{center}
REFERENCES
\end{center}

[1] A. Hocquenghen, Codes correcteurs d'erreurs, Cliffres 2, pp.
147-156(1959).\\

[2] R. C. Bose and D. K. Ray-Chaudhuri, On a class of
error-correcting
binary group codes, Inform and Control 3, pp.  68-79(1960).\\

[3] R. E. Blahut, Theory and practice of error-control codes, Addition-Wesley Publishing Company, 1984.\\

[4] T. B. Pei and C. Zukowski, High-Speed parallel CRC circuits in
VLSI, IEEE Trans. Commun., vol.40(1992), no.4 , pp.  653-657.\\

[5] R. J. Glaise, A two-step computation of cyclic redundancy code
CRC-32 fro ATM networks, IBM J,Res. Devel., vol. 41(1997), pp.
705-709.\\

[6] J. H. Derby, High-Speed CRC computation using state-space
transformation, in Proc. Global Telecommunications Conf., vol.1,
2001, pp. 166-170.\\

[7] K. K. Parhi, Eliminating the fanout bottleneck in parallel long
BCH encoders, IEEE Trans. Circuits and Systems--I:Regular Papers,
vol. 51(2004), no. 3, pp. 512-516.\\

[8] X. Zhang and K.K.Parhi, High-Speed architectures for parallel
long BCH encoders, IEEE Trans. VLSI systems, vol.13(2005), no.7, pp.
872-877.\\

[9] K. K. Parhi, VLSI digital signal processing systems, John
Wiley  Sons Inc., 1999.\\

[10] R. J. McEliece, The theory of information and coding: a
mathematical framework for communication,
Cambridge University Press, Cambridge, 1984.\\

[11] R. J. McEliece, Finite fields for computer scientists and engineers, Kluwer academic Publishers, 1987.\\

\end{document}